# Music, Immortality, and the Soul
Dean Rickles

> Such harmony is in immortal souls
> But whilst the muddy vesture of decay
> Doth grossly close it in, we cannot hear it.
>
> Lorenzo in *Merchant of Venice*

> Can there be a more spiritual delight than that which music transports us into?
> F. W. J. Schelling[1]

> Every music lover knows intuitively that music embodies a certain truth, but few go so far as to obey this intuition and search for truth by way of music.
>
> Joscelyn Godwin[2]

The great Viennese musicologist, Viktor Zuckerkandl, once wrote that "[M]usic is the temporal art *par excellence*" (1956, p. 151). In one sense this is of course quite true. Indeed, Goethe famously described architecture (at least that of the cathedral builders) as "frozen music," implying that music itself is essentially *unfrozen* or dynamical. Of course, one of the most basic elements of musical structure is its meter, that expresses, in this case, a temporal invariance that is neurophysiologically bound up with time processing and temporal phenomenology. What could be more obvious than this we might ask? Yet, at the same time, and as this paper will discuss, music is the *atemporal* art *par excellence*. Of course, I will have to explain what I mean by "atemporal" here. But I can say that what I do not mean is *spatial*; there is a sense in which music is neither in space nor time.[3] In this sense, unlike, say, film, which is also a temporal art, of necessity, film is *not* also an atemporal art.[4]

---

[1] *Clara, or, On Nature's Connection to the Sprit World*. State University of New York Press, 2002, p. 72

[2] *Cosmic Music: Musical Keys to the Interpretation of Reality*. Inner Traditions, 1989, p. 8.

[3] As Roger Scruton rightly points out, music is distinguished from mere sounds only by considering them "without focusing on their material causes" (2007, p. 229). Indeed, as I have argued in (Rickles, 2017), even the most basic temporal properties of music, such as meter, are not *in* the sound signal itself, but are largely imposed by the listener—Scruton (ibid.) notes that this fact was known even to Aristotle's student Aristoxenus of Tarentum. Hence, there are troubles, even at this level, with taking music to be nothing but its material realisation in sound.

[4] At a far stretch, one might point to the timelessness of certain *archetypal* elements appearing in films.

Music, then, somewhat like *information*, is 'realization independent.' However, as with information (such as the software running on some specific computer), there is a common opinion that without hardware the information (the software) has no kind of existence whatsoever. In the case of music this sometimes goes by the name "the trumpet argument." In a nutshell: no trumpet (or whatever instrument), no music! In more formal terms, music exhibits *instantiation dependence*.[5] It demands material realisation.

But, given this situation, we seem to be faced with a direct contradiction, *viz*.:

"[M]usic is the *atemporal* art *par excellence*" **and** "[M]usic is the temporal art *par excellence*"

The contradiction is, however, only apparent, and a result of viewing music from two possible perspectives. That it has these two perspectives is the focus of this paper. In particular, the way in which these two aspects of music allow it to function as a kind of conduit between transcendent and immanent; immaterial and material. This can help explain the power of music to touch places deep in the soul (the part of us that transcends matter and time), that other forms of art struggle to reach. It is perhaps no surprise that initiates into the so called *mysteries* consider music to be of special importance.

The harmony of the spheres was, of course, used to refer to the idea that the universe itself is engaged in the production of a kind of cosmic music, in perfect harmony,[6] with the lower levels (matter, humans, etc) re-enacting this in the best case—correspondences were, of course, also found between the number of distinct tones in an octave and the number of planets. Our subject matter leads us into issues that intersect with the notion of the harmony of the spheres. Indeed, according to the view I present, the layers of reality mirror one another, as microcosm (Earthly, material, temporal) to macrocosm (heavenly, immaterial, timeless). Giorgio Scalici (2019) has shown that the Wana people of Central Sulawesi in Indonesia view music in exactly these terms, as a cord between the human world and the occulted world of spirits. Indeed, it was used as a kind of existential therapy in a shamanic ritual (*momago*) designed to help a person find their soul, and so save their life from soul loss. It is viewed as a way of connecting the finite to the infinite, and employs it for positive transformative effects. Tommaso Palamidessi, founder of "Archeosofica" (which places special emphasis on the spiritual function of music), more recently, makes a very similar claim:

---

[5] One can find this viewpoint explicitly developed in physics in David Deutsch's 'constructor theory' approach to the laws of physics (2013). The idea is that there is substrate independence in one sense, namely in what *kind* of material some information is realised in, but substrate dependence in the sense of demanding *some* material basis.

[6] However, we will see that part of this larger, perfect harmony involves symmetry *breaking* as much as symmetry.

Based on our observations and experiments, hearing a musical piece is not only a sensorial or acoustic phenomenon. It is not only an aesthetic and ethical experimentation, because in the humans, equipped with ears, there is a psychosomatic and psycho-spiritual and then spiritual-divine process. It can become, following a particular methodology, the setting off to the relative unification with the uncaused Cause of the Created. It is a unification in a harmonic synthesis of the different aspects of the human personality (physical, emotive, mental, spiritual), both conscious or unconscious. It uses the numerous psychological techniques of liberation, guided and helped by the One who can do everything because He is the uncreated foundation of the creation of the Cosmos and of humans (Palamidessi n.d., cited in Coradetti & Cresti, 2022, p. 28).

Of course, even though modern people might no longer think in these terms, we still often and in great numbers turn to music when we require care for the soul, if nothing else as a way of remembering that we have such a thing. Our actions speak louder than words here.

1. *Infinity, Eternity, Finitude and Matter*

> Philosophy is an attempt to express the infinity of the universe in terms of the limitations of language
> (Whitehead, 1941, p. 14)

Whitehead's remark implies a reality that is forever veiled from us, at least using conceptual means. It presents philosophy as a way of probing an essentially unsolvable mystery, forced as we are into representing via finite concepts and categories. Music, by the same token, is an attempt to express the infinity of the universe in terms of the finite limitations of matter and radiation. Bringing music into material form, is akin to a symmetry breaking of the infinite and eternal, into time and creation. Stanley Cavell skirts the same kind of issue in his comparison of Schoenberg's serialism and Wittgenstein's view of music as akin to grammar:

> My suggestion is that the Schoenbergian idea of the row with its unforeseen yet pervasive consequences is a serviceable image of the Wittgensteinian idea of grammar and its elaboration of criteria of judgment, which shadow our expressions and which reveal pervasive yet unforeseen conditions of our existence, specifically in its illumination of our finite standing as one in which there is no complete vision of the possibilities of our understanding—no total revelation as it were—but in which the assumption of each of our assertions and retractions, in its specific manifestations in time and place, is to be worked through, discovering, so to speak, for each case its unconscious row. (Cavell, 2000: p. 182)

There is a sense in which music, in this form, mimics the act of creation, in which the cosmos is an essentially evolutionary entity that must *become*. A somewhat similar debate occurs in looking at mathematics from an ontological point of view. In particular the treatment of the real numbers. There are curious properties of real numbers that seem to put them, like music, in the realm of the

*transcendent*: in terms of the amount of information to specify them, one requires infinite computer time since there is no repeating pattern to their decimal expansions. One must simply evolve the sequence, working through it, despite the fact that it might have a perfectly situated home in Platonia. In other words, bringing them into this world demands a temporal element. It is an understanding of mathematics (known as intuitionism or constructivism) that must be "worked through," stage by stage, necessarily involving a temporal, step-wise process, rather than an eternal once-and-for-all structure (see Gisin, 2021). Through mathematics, like music, we get a sense of the infinite and eternal.

Philosophers of music have long discussed its curious ontological aspects. However, earlier thought from mystical thinkers went down similar paths and, I argue, much further. The most obvious of these paths is, of course, the theme of this book: that of Pythagoras and the harmony of the spheres, also involving the close linkage of mathematics and music. We can sum up the basic ideas of the Pythagorean school, originally associated with the idea of the harmony of the spheres that forms the theme of this book, as involving number being at the heart of reality, and music (as a form of number) having an effect on the human (mind, body, and soul). As the opening passage from the Merchant of Venice suggests, ordinary humans, clothed in their earthly, decaying bodies, subject to time, cannot hear the music of the spheres, which is a music of the soul: but music is nonetheless about as close as we can get to experiencing it. It draws us nearer to that which is immortal and immaterial while we think of ourselves as limited. On this, Max Heindel writes:

> None other ranks so high as the musician, which is reasonable when we consider that while the painter draws his inspiration chiefly from the world of color—the nearer Desire World—the musician attempts to bring us the atmosphere of our heavenly home world (where, as spirits, we are citizens), and to translate them into the sounds of earth life. His is the highest mission, because as a mode of expression for soul life, music reigns supreme. That music is different from and higher than all the other arts can be understood when we reflect that a statue or painting, when *once created, is permanent*. They are drawn from the Desire World and are therefore more easily crystallized, while *music, being of the Heaven World, is more elusive and must be re-created each time we hear it*. It cannot be *imprisoned*, as shown by the unsuccessful attempts to do so partially by means of such mechanical devices as phonographs and piano-players. (Heindel, 1922, p. 128, my emphasis)

Of course, "imprisoned" most strongly suggests gnosticism [associated more so with neo-Platonism]: music cannot be forced into time - it cannot ever be rendered *purely* material. When we perform music, we are bringing down the heavens/the macrocosm into the microcosm. So, while Michelangelo's David *just is* that piece of carved marble, and is itself confined or imprisoned; music needs to be brought *into* time, and is able to leave it. When we listen to a piece of music, we only hear a segment that is able to fit into the present moment of awareness, and we must integrate it.[7] We know the music itself can never be fully brought into being, and so we surmise it must be elsewhere, not in time. Similarly with mathematics, hence Pythagoras' original intuition.

A core idea of gnosticism, e.g. as espoused by Basilides and Valentinus, is that the material universe is inherently evil. The creator of the material world (and with it, time), the demiurge, the

---

[7] As Marvin Minsky asked: "How [this can] be, when there is so little of it present at each moment?" (1993, p. 338).

Old Testament God, is a deeply deluded deity on this account, forgetting that the monad (the true creative power) comes before all, including it. Knowing the true divine source, beyond this deity, is enough, on this account, to achieve liberation from matter. Heindel, like the gnostics, clearly possesses a distaste for the material world, which, while recognizing its obvious deficits, I do not share. I rather like the world, and I like music, and I think their interrelationship is vital for bringing life and spirit into matter. The fact that the composers were humans rather than angels is what provides power to the music. I think there is a better way of thinking about music's elevated status, rather than seeing it as breaking into a prison. Music exists ultimately as an element of the non-dual domain, which can leak out into the dual domain by taking form. Composers are able to access that domain, translating it into form, like demiurges in their own right. Like other kinds of subject matter, music is archetypal and symbolic at a very deep level—here I have in mind things like maths, sexual union, dreams, and so on. So I agree with Heindel that music in some sense lies beyond matter, but neither can it be divorced from the material beings it serves and whose perceptual faculties it is grounded in. Without this material grounding, in joy and pain, and the human condition, music would become sterile.[8]

Arnold Schoenberg, in his *Style and Idea,* made a similar remark (1984, Essay 2), directly referring to Gustav Mahler's symphonic progression in which Mahler himself ascended like a saint (*literally*, according to Schoenberg[9]): The 1st symphony is youthful, vital, and full of high spirits. Ditto the second, of which Schoenberg claims it caused "the violent throbbing of my heart." The 3rd he called "a thunderbolt." The 6th sounds a period of resignation. The 8th a leaving of the material world. And by the 9th symphony we witness a near-passionless and cold soundscape as viewed from the perspective of material existence and human life, a soundscape that is "most strange … no longer couched in the personal tone."[10] This Ascension involves an interesting feedback loop. Angels or saints cannot compose the music that can lead to angels and saints. The heavenly music is not for humans, as we know from Lorenzo, since "[s]uch harmony is in immortal souls…we cannot hear it." Humans require human music to lead them up the rungs of divinity. They need music that retains some sense of the heavenly harmony, but yet can move them through their physical attributes and emotions. Music can itself fall into matter, while retaining the cord to the timeless realm it also resides in. In this way it can transmit information, as a kind of remembrance, of a world outside space, time, and causality. In this sense it has much in common with synchronicities, in which a disruption occurs in the material world that indicates something other, something of a different order of existence. Often, these disruptions in the normal order are the only things that can serve to remind us, through a deep experience (knowing) that there exists another order of existence.

---

[8] Emotions themselves sit interestingly outside of matter and cognition, themselves acting as a kind of conduit between them. So it is not surprising to find links between music and the emotions—my thanks to Ange Weinrabe for drawing this feature to my attention.

[9] See Stuckenschmidt (1977, p. 103).

[10] I admit, I don't quite view the 9th symphony the way Schoenberg describes, and view it as much as stirring as the others. But the idea Schoenberg is getting it is quite correct, and it is true that the closing Adagissimo is otherworldly to the extent that we can easily imagine a conclave of crystalline, light beings listening along to it in silent stillness.

We can assume that Heindel would appreciate Mahler's 9th over the 1st. But we must not forget the role played by the 1st in the initial upward rising out of matter.[11] In this way, we must also not forget the necessity of matter, nor forget that it contains its own kind of divine role. We can see this kind of view at work, in which music is seen as bringing down the divine into matter, in the work of Louis Claude de Saint-Martin (1742-1803) to which we now turn. Central to this is the feature that makes the material world what it is: duality and polarity. It is the tempestuous strife of opposites that puts the fire in the earlier works of Mahler, and their relative absence that renders the 9th so austere.

*3. Thoughts on music of the "Unknown Philosopher"*

> "No self-respecting Demiurge would create a universe out of concord alone."
> Joscelyn Godwin[12]

Louis Claude de Saint-Martin, associated with the School of Martinism, was strongly connected to the study of the deepest mysteries. Indeed, a pseudonym of his was "lover of secret things"—he was also the first to translate the writings of another mystic Jakob Böhme from German into French. Martinism is a kind of esoteric Christian mysticism concerning itself with broadly gnostic themes: the fall of man, the yearning that is felt from the privation of the divine source, and the process of his return or ascension, which is a kind of reintegration into source or Oneness. Music played a key role in his thinking on the subject—he was himself an accomplished violinist.

He distinguishes between two kinds of music:

I. "Practical music," is the everyday music: it is temporal; exoteric; quotidian.
II. "Natural music" is Platonic, timeless, esoteric.[13]

As with gnosticism: practical music can only approximate natural, Platonic music, and indeed that is its purpose: to bring down the celestial into the earth. Saint-Martin employed the Seal of

---

[11] We will look at Schoenberg's musical adoption of this idea, in his opera *Moses and Aaron* and his unfinished *Die Jakobsleiter*, in a later section.

[12] Joscelyn Godwin, 1991, p. 87.

[13] We might also speak in terms of the substrate independence, mentioned earlier, for this form: there is no dependence on material instantiation (hence Platonic).

Solomon, with its interlocking triangles pointing upwards and downwards to represent this role, with practical pointing down to Earth and natural pointing upwards to the heavens, and their entanglement representing the equal importance of both. The composer's task is to aim for the Platonic/natural, rather than the earthly, while being inescapably embodied.

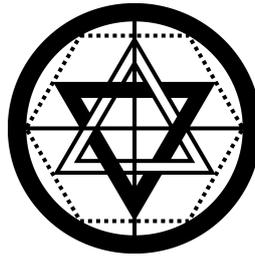

Saint-Martin also viewed the strife of opposites in music as analogous to the problem of good versus evil. He viewed the minor key as a form of evil, with the perfect octave damaged by the 7th. Yet, the analogy goes deeper for Saint-Martin since "[w]e know, too, that this chord of the seventh tires the ear, holds it in suspense, and demands to be *saved*" (Godwin ed., 1995, p. 22). Saint-Martin draws consequences for the immortality of the soul from this, pointing out how our own battle in duality is simply the preamble to a perfect cadence:

> "One can see…that just after this musical cadence one necessarily returns to the common chord which restores all to peace and order, it is certain that after the crisis of the elements, the Principles which have fought over them will also regain their tranquility. And applying the same to man, one just sees how *the true knowledge of music might preserve him from fear of death*: for this death is only the trill which ends his state of confusion, and restores him to his four consonances". (Godwin ed., 1995, ibid.)

There is something truly cosmic about this. Normal music is temporal, and so it ceases, just as it commences. The principle music will never cease, since it resides outside of time. In this way music becomes a matter of life and death, and immortality. It is useful to compare this with Leibniz's ideas, since we have a tale of two distinct theodicies here. For Leibniz, in his *The Ultimate Origin of Things,* what he calls "the law of delight comes from this disharmony:

> "Someone who hasn't tasted bitter things doesn't deserve sweet things, and indeed won't appreciate them! This is a *law of delight*: Pleasure doesn't come from uniformity, which creates disgust and makes us numb rather than happy." (G. W. Leibniz, in Bennett trans, p. 6)

While it seems superficially as though Saint-Martin is opposed to disharmony, like Leibniz he clearly sees the necessity for it, writing, of music, that "far from causing the least fault in it, are its nourishment and its life" (Saint-Martin, ibid.). Music is nourished by dissonance. And, for both Leibniz and Saint-Martin, this is part of a larger cosmic story in which dissonance is the very stuff of life. For Leibniz, the apparent asymmetry is part of a larger harmony aiming at an overall perfection of its parts. We can easily see how this can be represented in modern physical theory, where it appears in the context of the theory of symmetry breaking, which takes place between various kinds of ordered state.

*4. Chaos, Order, and Symmetry Breaking*

We do not appreciate disharmony in architecture or in the bodily, human form in quite the same way as Leibniz and Saint-Martin suggest. Rather, we feel a kind of visual pain from such ugliness. However, here we might still consider other buildings and other human forms to be the appropriate contrast class, providing the friction or polarity required to see something as beautiful and orderly (and perhaps even more so, given the contrast cases). Music is special here, then, since it contains its own essential disharmony, rather than requiring it by relationship to another work. An overly orderly musical work is bland, even if heard sandwiched between a cacophony. This is yet another aspect of the essentially temporal (yet atemporal) features of music. It exemplifies the close link to deeper cosmic issues, as mentioned above.

It is clear, that if we are in some ordered, harmonious state, then we must break symmetries to get a new order. When this order establishes itself (high entropy), then a new breaking is required to keep the system in activity (or low entropy). There would be no life without this repeated symmetry breaking, avoiding equilibrium. Like life, good music contains phases far from equilibrium. The very direction of time itself is a gift that comes from an entropy gradient[14] where we fortunately find ourselves, locally at least, with an absence of uniformity of energy (an asymmetry, relative to the cosmos as a whole). According to particle cosmology, symmetry breaking in the early Universe was responsible for the material structure we find. In the earliest phases, high pressures and temperatures prevented the creation of elementary particles, so there could be no formation of stable objects until the Universe had cooled down, breaking one symmetry into another. Such symmetry breaking processes are often modelled with a so-called Mexican Hat potential, in which the system starts like a ball on the top (a symmetrical state, as you can see), but an unstable one that demands resolution, ball falling to a lower state. This corresponds very closely to the kinds of tension that are built up through dissonance, that demands resolution, momentarily moving back into order, only to build back into tension, as gives motion and energy to all material things.

---

[14] In other words, there is low entropy (i.e. a highly ordered state) in the past, which progressively increases, since it is more probable at each time step to move into a state of lower order. Life seems to be capable of reversing this degradation, injecting order into systems that would otherwise tend towards stagnation.

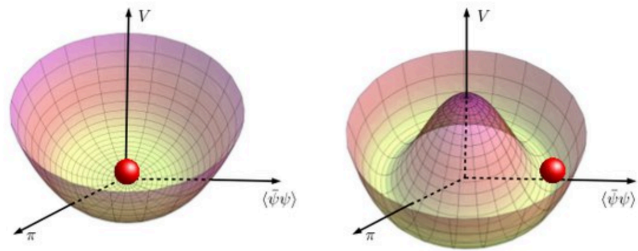

But the development of music itself, as a historical entity, like a culture, can also be seen through this lens. Though humans like order, and need a large degree of it, there is also the need to stave off equilibrium/stagnation in our activities and cultural products. Schoenberg viewed his own role exactly along these lines: as a necessary symmetry breaker, to allow for something new to come. He referred in this regard to his system of dodecaphony (i.e. the 12 tone approach using an ordered series of all twelve chromatic tones as the basis for a musical work) as the "law of emancipation of dissonance," noting that "[a] style based on this premise treats dissonances like consonances and renounces a tonal center." This required radical action since the traditional diatonic order (e.g. whole steps and half steps) had persisted since Ancient Greek times and was thoroughly entrenched. Schoenberg had to fill himself with the feeling the divine creative force in this battle, since something genuinely new was required to break a frozen, crystallised musical order (itself part of an equally frozen cultural order):

> To understand the very nature of creation one must acknowledge that there was no light before the Lord said: 'Let there be Light'. And since there was not yet light, the Lord's omniscience embraced a vision of it which only His omnipotence could call forth … A creator has a vision of something which has not existed before this vision. And a creator has the power to bring his vision to life, the power to realise it. (Schoenberg, 1984, pp. 214-15).

Schoenberg had a list of divine creators: Mahler, Wagner, Beethoven, Mozart and Bach. They were needed when some form/age had exhausted itself - in his case tonality, and in previous cases including other elements, such as form. Part of the symmetry breaking process is a dip into chaos, since what was needs to be dissolved, and what is can not yet be. So one has a state of pure chaos which contains within it all potentialities. Only here can the new appear (see Rickles, 2024 for a detailed explanation why).

But Schoenberg exhibits a viewpoint similar to that of Heindel, which exhibits a high degree of the (necessary) Luciferian push towards spirit and new heights. However, while this push is needed to make a strong enough shift, if left unchecked one is led to the cold and austere Mahler (which, unsurprisingly, Schoenberg approved of[15]). One can see this exuberant upwards vision here:

> As with all ancient peoples, it is our destiny to spiritualize ourselves, to set ourselves free from all that is material…We want to perfect ourselves spiritually: we want to be free to dream our dream of God—as all ancient people, who have left materiality behind them. (ibid.)

Yet music that is too elevated cannot serve this function. Too much spirit and not enough material and one no longer has music fit for humans, and nothing to lift them up with.[16] Hence, the purpose of music, whether by design or not, is to reconcile matter (limitation/finitude) and spirit (infinity). They point to a perfect method for straddling two worlds. Hence, this idea pushes against the various orthodox stances, especially in the various religious frameworks, according to which existence involves a battle between a pair of polar opposites, with one side the correct and the other to be demolished. Thus, while Schoenberg might be correct in stating that "music gives humanity an immortal soul," we might supplement this with Heinrich Schenker's remark that "music mirrors the human soul." While Schoenberg looks to the heavens, Schenker redirects the light downwards[17]:

> The genius gathers the gazes of men unto himself; woven out of these gazes directed upward to the genius there arises, as it were, a mysterious cone of light, the most inspiring symbol of a great community of mankind. Without such a cone of light, the mass of mankind remains in a plane that extends in all directions hopelessly, desolately, to infinity. (Schenker, 1997, 3:105/69)

This is a symmetry breaking description, but one that retains a love for humanity, even in its lower condition. The future of humanity cannot be an either materialist or spiritualist one (this is the source of many problems); rather, it must bring them together, since each alone leads to a different kind of chaos which is avoided only through a balancing/rebalancing act. Contrast this with Schoenberg:

> Present and genius have nothing to do with one another. The genius is our future. So shall we too be one day, when we have fought our way through. The genius lights the way, and we strive to follow. Where he is, the light is already bright; but we cannot endure this brightness. We are blinded, and see only a reality which is as yet no reality, which is only the present. But a higher reality is lasting, and the present passes away. The future is eternal, and therefore the higher reality, the reality of our immortal soul, exists only in the future. There is only one content, which all great men wish to express: the longing of mankind for its future form, for an immortal soul, for dissolution into the universe – the longing of this soul for its God … And this longing is transmitted with its full intensity from the predecessor to the successor, and the successor continues not only the content but also the intensity, adding proportionally to his heritage. This heritage carries responsibility, but it is imposed only upon one who can assume this responsibility. (Schoenberg, 1935, p.)

---

[15] Karlheinz Stockhausen was well aware of this process, and writing of his work *Lucifer's Dream*, notes that "It's nothingness, emptiness, he wants". He has Lucifer sing of the "Compression of the figures of human music" which amounts to a mixture of animal and angelic (1989, p. 106). What is required is neither Luciferic (with its push away from individualised matter and into perfect order) nor its opposite, Michaelic; but instead a treaty between the two forces, which Saint-Martin's seal nicely encapsulates.

[16] And, to quote David Temperley: "Why would the great composers have bothered to create such elaborate mental structures if they thought that these structures would never be shared by listeners?" (2011, p. 147).

[17] On the comparative study of Schoenberg and Schenker's spiritual approaches, see Arndt (2018).

What is "dissolution into the universe" but maximal equilibrium (i.e. maximal entropy)? What is this but an absence of life and creation? We might say *peace*, but, peace is not in equilibrium; that is only death. Just as a world directed only towards making all material is also death.

It is interesting to compare with the christian mystic, Jakob Böhme's cosmology here. He too places a heavy responsibility upon humanity. Indeed, without the reflecting capacity of some kind of conscious awareness, the entire universe of the creation would disappear into chaos (in the sense of no-thingness, no division, no opposites, equilibrium, perfect symmetry). The manifest universe ends when this happens, since there is nothing for it to be manifest too. Creation (cosmogensis) is, for Böhme, the divine's own mission to become whole; humanity participates in this mission through being a faithful mirror. The light that music shines down acts as food for us, that we can join in this participation.

If we buy this deep understanding music as a path to the soul, then we see all the more clearly the truth in Shakespeare's Lorenzo (Merchant of Venice, 5.1.89-94):

> The man that hath no music in himself,
> Nor is not moved with concord of sweet sounds,
> Is fit for treasons, stratagems and spoils;
> The motions of his spirit are dull as night
> And his affections dark as Erebus:
> Let no such man be trusted.

Music lines up matter wth soul and spirit, it is the product of union between these two realms. At the same time, in doing so, it manages to consistently manage the One and the Many.

*5. Dual-Aspect Monism and Music*

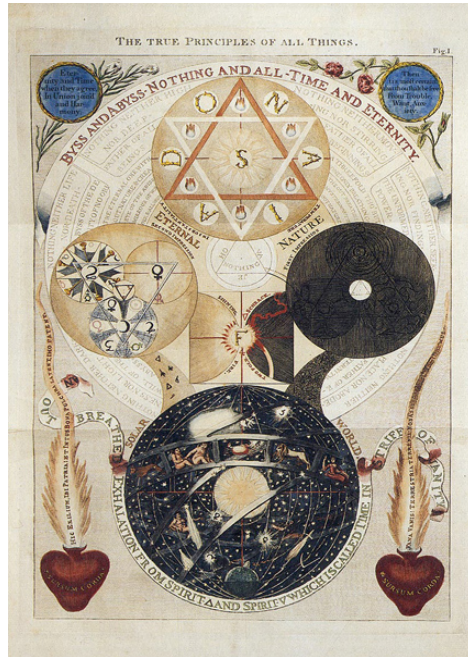

> "Of all artistic media, music is perhaps most at ease in demonstrating the collapse of dualities."
>
> Daniel Albright, *Music's Monisms*

In this final section, I very briefly describe a newly developed framework (decompositional dual-aspect monism[18]) that might serve to explain how music performs the function of linking the temporal (matter and duality) to something atemporal (higher and non-dual). It shows how both the curious temporal yet atemporal nature of music can mimic the idea of a soul and its incarnation into matter which is correlated with a mind. Music becomes *alive* in some performance, but *it* doesn't really go anywhere. Its source is beyond the performance and beyond space and time. We can view each performance as a kind of reincarnation. To paraphrase Jakob Böhme, music is where "time and eternity meet" (see the figure below, from Dionysius Andreas Freher, designed for Böhme's works).

The basic idea of the approach is that one needs two polar aspects to *manifest* a universe, both a subject and an object (or probe and probed). But these cannot co-explain one another in a causal fashion, nor do we have very satisfactory ways of explaining why there is the correlation between mental and material, or whatever probe and probed we employ. The dual-aspect approach argues that they are two different projections of the same neutral source, which pick up correlations as a result of this common home. The *decompositional* approach starts with an undivided totality (in which nothing is yet manifest, and one has neutrality with respect to mind and matter) that is then

---

[18] The theory is fully developed in Atmanspacher & Rickles (2022).

split into pairs that some temporal) totality below).[19]

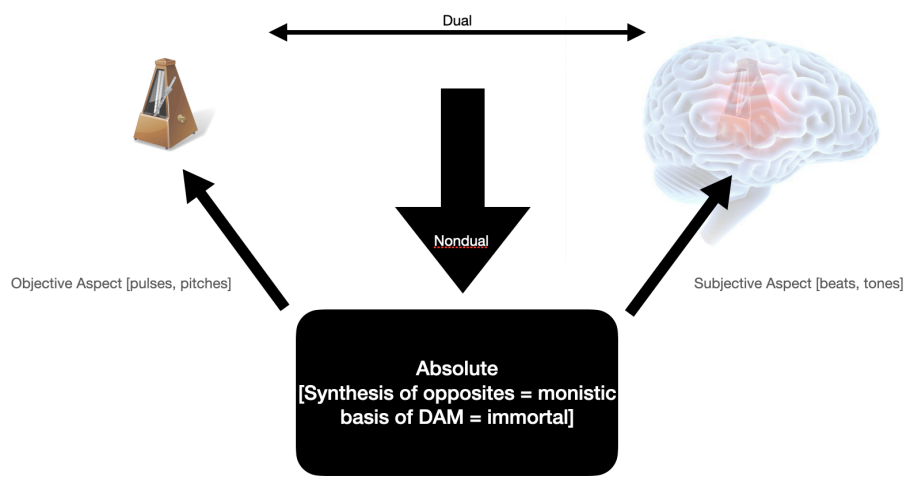

subject/object manifest (finite, local, aspect of the (see the figure

This has much in common with David Bohm's holographic theory of reality according to which what is fundamental is "neither mind nor body but rather a yet higher-dimensional actuality, which is their common ground and which is of a nature beyond both" (Bohm, 1995, p. 209). Hence, there is an origin of mind and matter in something else which does share their properties but has the *potential* for exhibiting those properties. Bohm gives a useful example of what this might look like. Imagine two cameras located on different sides of an aquarium in which there is one fish, one facing the front of the fish, one the side. Then imagine that each camera is linked to a screen, so that we only have access to this pair of screens, side by side. We do not know that there is one fish, and it would appear to be distinct fish because of the differing perspectives. We would see on one screen a fish coming towards us directly, while on the other a fish moving across the screen from right to left say. As the fish moves in different directions, we would see correlations between their movements. We might be led to postulate some causal connection between the 'two' fish (here standing for matter and consciousness). But we know in this case that would be a mistake: it is a single entity, viewed from different perspectives. Likewise, as Bohm says: "we do not say that mind and body causally affect each other, but rather that the [apparently correlated] movements of both are the outcome of related projections of a common higher-dimensional ground" (ibid.).[20]

---

[19] This has much in common with Spinoza's approach, and contrasts with most modern dual-aspect theories which assume that the mind-matter aspects are built up (composed rather than decomposed) from neutral elements.

Likewise, the 'incarnation' of music into the world involves some realisation in sound-generating systems (objects which generates pitches and beats) as well as a perceiving subject (which involves tones and accents). These give us the "apparently correlated" 'outside' and 'inside' views. Usually, we consider this split between subject and object as permanent and fundamental: the subject *picks up* information about the objects out there in the world, like a receiver—Scruton distinguishes between "the intentional object of a listener's experience" and "the material organization of sounds" (1999, p. 415). However, with music this simple division seems to come under pressure[21], and the situation is rather more like Hermetic philosophy would have it: in the words of William Blake, as a man is, so he sees (and hears). Music is the perfect vehicle for understanding Hermeticism: the outer and the inner are co-constitutive. This can be seen most clearly, I think, with the case of meter. There is a tension in this case over whether meter is in the head or in the world. Victor Zuckerkandl isolates the issues here, pointing out that "[I]t is not a differentiation of accents which produces meter, it is meter which produces a differentiation of accents" (1956, p. 169). If it is not fully part of the sound object, nor fully part of the experience, yet part of the music, then we must look beyond both. Zuckerkandl himself is well aware of the significance of this curious tension, writing:

> Those who believe that music provides a source of knowledge of the inner world are certainly not wrong. But the deeper teaching of music concerns the nature not of 'psyche' but of 'cosmos.' The teachers of antiquity who spoke of the music of the spheres, of the cosmos as a musical order, knew this (Zuckerkandl, 1956, p. 147).

This takes us back to the idea of a kind of feedback loop, in which the cosmos generates its manifestation, which leads back to union, especially in the case of musical manifestation which bears the imprint of its divine origin more clearly. Goethe expresses the concept thus:

> What appears in the world must divide if it is to appear at all. What has been divided seeks itself again, can return to itself and reunite (Goethe, 2016, p. 951).

Goethe is expressing here something very similar to the Sufi notion of "The Breath of the Merciful," understood as the cosmic dynamics that revitalises the Many in the One, and indeed decompositional dual-aspect monism is very similar (see the figure below). The link can be seen even more explicitly in the following:

> To divide the united, to unite the divided, is the life of nature; this is the eternal systole and diastole, the eternal *syncrisis* and *diacrisis*, the inhaling and exhaling of the world, in which we live, move, and have our being. (Goethe, 2014*, ¶739)*

---

[20] C. J. Jung argued similarly that the fact that "causal connections exist between the psyche and the body … point to their underlying unitary nature" (1971, p. 767).

[21] There are other more idealistic alternatives in which this direction is simply reversed, so that music becomes a mental construction—e.g. in Lerdahl and Jackendoff's "Generative Theory," in which they situate music theory as a branch of theoretical psychology: "Insofar as one wishes to ascribe some sort of 'reality' to these kinds of [musical] structure, one must ultimately treat them as mental products imposed on or inferred from the physical signal" (1983, p. 2).

By splitting, the One (the totality) generates (breathes out) individuality (though while always remaining an harmonic totality in some sense), which itself allows for relationship between the many parts, without which such relations as love would not exist. This splitting allows for states of knowledge too, since it generates knowers and known, which, according to Sufi thought, is how the Hidden Treasure could come to know itself. But the state of knowing is always a state of separation, in which a veil is placed between knower and known, and so the inhalation restores unity once again, which is a kind of love of a different order. This cyclic approach is exactly the kind of system needed to resolve the Luciferic/Michaelic tension, to bring out a world in which the forces towards and away from order are balanced, yielding a true harmony of the world.